\documentclass[twocolumn,english,aps,pra,superscriptaddress]{revtex4-1}
\usepackage[latin9]{inputenc}
\usepackage{color}
\usepackage{babel}
\usepackage{amsmath}
\usepackage{amssymb}
\usepackage{graphicx}
\usepackage[unicode=true,pdfusetitle,
 bookmarks=true,bookmarksnumbered=false,bookmarksopen=false,
 breaklinks=false,pdfborder={0 0 1},backref=false,colorlinks=true]
 {hyperref}
\hypersetup{
 urlcolor=blue, citecolor=blue, hyperfootnotes=blue, linkcolor=blue}
\usepackage{breakurl}

\makeatletter
\usepackage{mathrsfs}
\usepackage{epsfig}
\usepackage{amsfonts}
\usepackage[figuresright]{rotating}
\usepackage{dcolumn}
\usepackage{bm}
\usepackage{color}

\makeatother

\begin{document}

\title{Signatures of a quantum phase transition on a single-mode bosonic
model}

\author{Emmanouil Grigoriou}

\affiliation{Wilczek Quantum Center, School of Physics and Astronomy, Shanghai
Jiao Tong University, Shanghai 200240, China}

\author{Carlos Navarrete-Benlloch}
\email{corresponding author; derekkorg@gmail.com}

\affiliation{Wilczek Quantum Center, School of Physics and Astronomy, Shanghai
Jiao Tong University, Shanghai 200240, China}

\affiliation{Shanghai Research Center for Quantum Sciences, Shanghai 201315, China}
\begin{abstract}
Equilibrium phase transitions usually emerge from the microscopic
behavior of many-body systems and are associated to interesting phenomena
such as the generation of long-range order and spontaneous symmetry
breaking. They can be defined through the non-analytic behavior of
thermodynamic potentials in the thermodynamic limit. This limit is
obtained when the number of available configurations of the system
approaches infinity, which is conventionally associated to spatially-extended
systems formed by an infinite number of degrees of freedom (infinite
number of particles or modes). Taking previous ideas to the extreme,
we argue that such a limit can be defined even in non-extended systems,
providing a specific example in the simplest form of a single-mode
bosonic Hamiltonian. In contrast to previous non-extended models,
the simplicity of our model allows us to find approximate analytical
expressions that can be confronted with precise numerical simulations
in all the parameter space, particularly as close to the thermodynamic
limit as we want. We are thus able to show that the system undergoes
a change displaying all the characteristics of a second-order phase
transition as a function of a control parameter. We derive critical
exponents and scaling laws revealing the universality class of the
model, which coincide with that of more elaborate non-extended models
such as the quantum Rabi or Lipkin-Meshkov-Glick models. Analyzing
our model, we are also able to offer insights into the features of
this type of phase transitions, by showing that the thermodynamic
and classical limits coincide. In other words, quantum fluctuations
must be tamed in order for the system to undergo a true phase transition.
\end{abstract}
\maketitle

\section{Introduction}

Depending on the environmental conditions, systems may exhibit strikingly
different behaviors. To account for these differences, the concept
of `phase' is often employed \citep{greiner1995thermodynamics}. They
may be familiar equilibrium phases such as solid or liquid, but also
more exotic such as dynamical \citep{Heyl18} or topological \citep{Wen17}.
As one parameter is varied, the system can change phase through a
crossover or by undergoing a phase transition. In the presence of
the latter, the boundary between phases can be defined through a critical
point, around which the system will exhibit characteristic phenomena
such as the divergence of the correlation length and the slow down
of the dynamics, which are notorious hurdles to simulations. These
phenomena, often dubbed critical phenomena, come with some subtleties
that are still a subject of research. For example, the generation
of long-range order and the corresponding divergence of the correlation
length are still considered fundamental components of criticality,
although recent works \citep{MyungJoongJC,MyungJoongRabi} have hinted
at the presence of phase transitions in spatially non-extended systems.

An interesting feature of phase transitions is the fact that similar
critical behavior is observed across a broad spectrum of models coming
from seemingly unrelated topics in physics, chemistry, and even biology.
In the case of equilibrium continuous phase transitions, these connections
are now understood through the well-established concept of universality
\citep{sachdev2011quantum,KadanoffBook,GoldenfeldBook} and the emergent
nature of the macroscopic properties characterizing the equilibrium
phases. This results in the possibility of gathering critical behavior
in universality classes, characterized by how the emergent properties
behave around the critical point as the thermodynamic limit is approached.
As a natural consequence, in order to disentangle the essential physics
relevant to criticality from other intricate phenomena or to test
new theoretical and numerical tools, finding the simplest model within
a particular universality class is a very relevant task. 

This point is well illustrated within the realm of equilibrium quantum
phase transitions. Conceptually, these are a consequence of abrupt
changes in the ground state of the system as we smoothly vary a parameter
of its underlying Hamiltonian. The phase change is not driven by thermal
fluctuations and can occur even at zero temperature \citep{Vojta_2003}.
Consequently, it is sometimes said that the phase change is driven
by quantum fluctuations. Experimentally accessible imprints commonly
survive finite temperatures as indicated by, e.g., the superconducting-insulator
transition \citep{conductor-insulatorQPT} or the tranverse-field
Ising model \citep{dutta2015TransverseQPT}. In this respect, the
last decades have seen a flourishing number of technological platforms
leading the field of many-body physics to an ever increasing number
of experimental observations, in particular within the scope of condensed-matter
systems. In order to meet modern technological needs, materials with
complex electronic structures displaying conductor-insulator transitions,
heavy fermion compounds, and two-dimensional electron gases \citep{carr2011understandingQPT}
have become topics of vivid interest. Many conceptual mysteries remain
within such complex systems, with predictions heavily relying on costly
numerical methods \citep{vojta2007computing} and quantum simulating
platforms such as optical lattices \citep{Jaksch_2005,Bloch08,Bloch12,Dutta15}.
This complexity has highlighted the necessity for new theoretical
ideas and novel approaches, driving the community towards simpler
platforms. For instance, engineering of quantum phase transitions
in artificial nanoscale devices such as quantum dots \citep{QuantumDot,Rau13}
is one of such attempts.

Central to the mathematical framework of phase transitions lies the
concept of thermodynamic limit. This limit is obtained when the number
of available configurations of the system approaches infinity, which
traditionally has been associated with a divergent number of system
constituents \citep{sachdev2011quantum,Vojta_2003}, in turn usually
linked to the physical extension of the system approaching infinity.
This is indeed the case of the paradigmatic transverse-field Ising
model \citep{SuzukiBookIsing}, for example, where the thermodynamic
limit is obtained when the size of the lattice that hosts one spin
at each node diverges. Recently, however, the intriguing possibility
of using the infinite-dimensional Hilbert space of a single harmonic
oscillator (which in principle already provides an infinite number
of available configurations), has motivated researchers to study non-extended
systems such as the Dicke \citep{Bakemeier12}, quantum Rabi \citep{MyungJoongRabi},
or Jaynes-Cummings \citep{MyungJoongJC} models. All of these consist
of a single bosonic mode coupled to a finite spin. It is argued that
its ground-state energy possesses a critical point as a function of
the coupling strength, with the thermodynamic limit determined by
the ratio between the characteristic energy scales of the bosonic
mode and the spin. Moreover, for the first two models, the phase transition
is found to be in the same universality class as previously known
mean-field models such as the Lipkin-Meshkov-Glick (LMG) model \citep{LMG65,Ribeiro08},
where an infinite number of two-level systems interact all with one
another, which can be equivalently formulated as a model for a single
large spin.

In this work, we further argue that phase transitions can be defined
in systems that are not spatially extended, by going to the extreme
of developing a model containing only a single bosonic mode. According
to conventional definitions, the system undergoes a change as a function
of a control parameter that displays all the characteristics of a
second-order phase transition. In particular, to our knowledge, we
provide here the simplest model within the same universality class
as the quantum Rabi and LMG models. The simplicity of the model allows
us to characterize it from first principles in all the parameter space,
giving solid support to the various analytical approximations that
we use to get physical insight. Furthermore, our model allows us to
prove an interesting point: the thermodynamic limit in which the critical
behavior appears scales with the number of excitations and is shown
to coincide with the classical limit in which quantum fluctuations
become negligible. The simplicity of our model also makes it a perfect
one as a building block for more complex lattice structures that might
present interesting interplays between many-body and local critical
phenomena. 

Let us outline the contents of this article. In Section II we introduce
the model, its approximate ground states, and the numerical approach
that allows us to study it in all parameter space. In Section III
we present our main results, studying the behavior of the system around
the critical point and determining that it offers all the signatures
of a second-order phase transition. In Section IV we discuss the sensitivity
of the model to symmetry-breaking perturbations when in the ordered
phase, and in Section V we conclude and discuss some subtleties related
to the equivalence between the thermodynamic and classical limits.

\section{Model and method\label{sec:Model-and-method}}

Consider the single-mode bosonic Hamiltonian

\begin{align}
\hat{H} & =\hat{a}^{\dagger}\hat{a}-\frac{\varepsilon}{2}\left(\hat{a}^{\dagger2}+\hat{a}^{2}\right)+\frac{1}{2L}\hat{a}^{\dagger2}\hat{a}^{2},\label{H}
\end{align}
where $\hat{a}^{\dagger}$ and $\hat{a}$ are creation and annihilation
operators satisfying the canonical commutation relation $[\hat{a},\hat{a}^{\dagger}]=1$.
Note that we have normalized the energy scale to the parameter of
the first term, which depending on the implementation has different
physical significance (e.g., the detuning with respect to a driving
field in quantum optics \citep{CNB-QOnotes} or a chemical potential
in an atomic gas \citep{Jaksch_2005,Bloch08,Bloch12,Dutta15}\textbf{)}.
The parameter $\varepsilon>0$ breaks particle-number conservation
and is associated with the coherent injection of pairs of bosons,
while the parameter $1/L>0$ is associated to the nonlinearity or
repulsive interactions between the bosons. We will see that $\varepsilon$
is responsible for crossing a phase transition, while $L$ controls
how close we are to the thermodynamic limit. Note that this Hamiltonian
has a discrete $Z_{2}$ symmetry, as it is invariant under the parity
transformation $\hat{U}=e^{\mathrm{i}\pi\hat{a}^{\dagger}\hat{a}}$,
which acts as $\hat{U}^{\dagger}\hat{a}\hat{U}=-\hat{a}$.

One motivation to study this model is that, in contrast to the previous
ones, we find ways to analyze it in all the region of the parameter
space $(\varepsilon,L)$ both through approximate analytical techniques
and first-principles numerical ones. In the reminder of this section
we explain the different approaches that we use: Coherent-state (classical)
ansatz, Bogoliubov-de Gennes theory around the classical minima, general
Gaussian ansatz, and full numerical simulations via a well-chosen
basis of the Hilbert space. These methods will univocally show that
there is a second-order phase transition at $\varepsilon=1$ for $L\rightarrow\infty$
that belongs to the same universality class as the phase transition
present in the quantum Rabi and LMG models.

\begin{figure*}[t]
\includegraphics[width=1\textwidth]{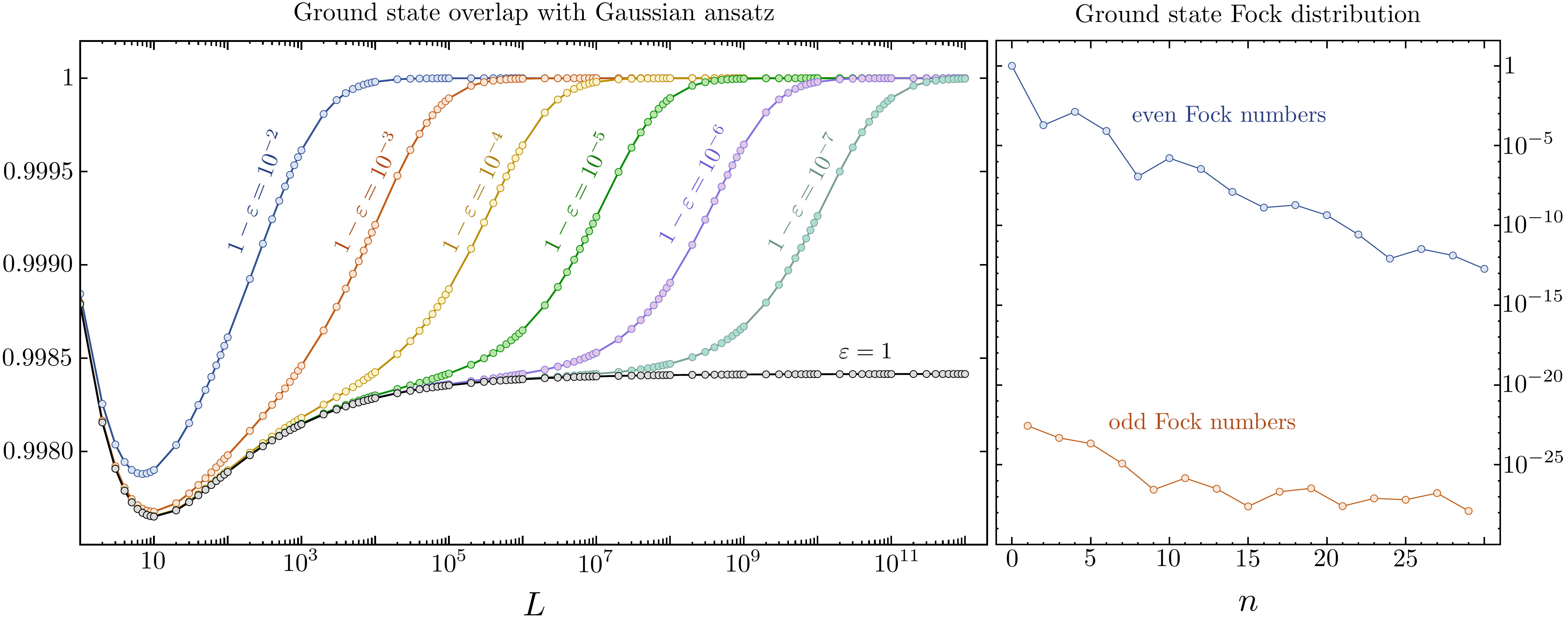}\caption{(Left panel) Overlap $|\langle0|\hat{S}^{\dagger}(\bar{z})|\psi_{\text{GS}}\rangle|^{2}$
between the true ground state $|\psi_{\text{GS}}\rangle$ of the system
and the Gaussian ansatz $\hat{S}(\bar{z})|0\rangle$ as a function
of the system size $L$ for different values of the pair injection
rate $\varepsilon$ close to the critical point $\varepsilon=1$.
(Right panel) Fock distribution $|\langle n|\hat{S}^{\dagger}(\bar{z})|\psi_{\text{GS}}\rangle|^{2}$
of the true ground state of the system for $L=10^{12}$ at the critical
point. \label{Fig_GaussianOverlapFockDist}}
\end{figure*}

Let us first explore the classical limit of this model. We make a
coherent-state ansatz $\vert\alpha\rangle$ \citep{CNB-QOnotes},
characterized by being a right (left) eigenstate of the annihilation
(creation) operators, i.e. $\hat{a}\vert\alpha\rangle=\alpha\vert\alpha\rangle$.
The corresponding classical energy function is
\begin{align}
E(\alpha,\alpha^{*}) & =\langle\alpha\vert\hat{H}\vert\alpha\rangle=\frac{|\alpha|^{4}}{2L}+|\alpha|^{2}-\frac{\varepsilon}{2}(\alpha^{*2}+\alpha^{2}).\label{eq:ComplexELand}
\end{align}
Since the first two terms are positive and depend only on the magnitude
of $\alpha$, the energy is obviously minimized for $\alpha\in\mathbb{R}$,
since then the last term is the smallest possible for any given magnitude
$|\alpha|$. The classical energy takes the simple form
\begin{equation}
E(\alpha)=(1-\varepsilon)\alpha^{2}+\frac{1}{2L}\alpha^{4},
\end{equation}
which changes from a single-well structure for $\varepsilon\leq1$
with minimum at $\alpha=0$, to a double-well one for $\varepsilon>1$
with minima at $\alpha=\pm\sqrt{L(\varepsilon-1)}$. In other words,
the trivial state $\alpha=0$ becomes unstable for $\varepsilon>1$
in favor of two non-trivial states each of which break the $Z_{2}$
symmetry. As we will see later, quantum mechanically it is shown that
there is indeed a second-order phase transition. This simple classical
picture lays the intuition of the system. 

In order to go further quantum mechanically, but still allowing for
some analytics, we consider small quantum fluctuations around the
classical minima $\alpha$. To this aim, we move to a picture displaced
to the corresponding phase-space location, defined by the unitary
transformation operator $\hat{D}(\alpha)=\exp(\alpha\hat{a}^{\dagger}-\alpha^{*}\hat{a})$,
the so-called displacement operator. Any state $|\psi\rangle$ is
transformed into $\hat{D}^{\dagger}(\alpha)|\psi\rangle$, which evolves
according to the Hamiltonian $\hat{h}=\hat{D}^{\dagger}(\alpha)\hat{H}\hat{D}(\alpha)$.
Using $\hat{D}^{\dagger}(\alpha)\hat{a}\hat{D}(\alpha)=\hat{a}+\alpha$
and truncating to second order in creation and annihilation operators
(which in this picture correspond to fluctuations around $\alpha$)
reads
\begin{align}
\hat{h}\approx E(\alpha)+\Delta(\alpha)\hat{a}^{\dagger}\hat{a}-\frac{\sigma(\alpha)}{2}\left(\hat{a}^{2}+\hat{a}^{\dagger2}\right),\label{h}
\end{align}
with\begin{subequations}\label{Deltasigma}
\begin{align}
\Delta(\alpha) & =1+\frac{2\alpha^{2}}{L},\\
\sigma(\alpha) & =\varepsilon-\frac{\alpha^{2}}{L}.
\end{align}
\end{subequations}The linear term vanishes because the corresponding
coefficient is equal to $\partial E/\partial\alpha$, which vanishes
at the classical minima. We expect higher-order corrections to vanish
in the $L\rightarrow\infty$ limit, as we indeed show when discussin
Fig. \ref{Fig_GaussianOverlapFockDist} and the Gaussian ansatz below.
Note that the eigenstates of $\hat{h}$ are independent of $L$, since
expressions (\ref{Deltasigma}) are independent of $L$ at any of
the classical minima $\alpha$. Therefore, the ground state coming
from this approximation can only be exact in the limit $L\rightarrow\infty$,
which we identify later with the thermodynamic limit. Below we discuss
how to approach the finite-$L$ case.

Being quadratic, this approximate Hamiltonian $\hat{h}$ is easily
diagonalized via a Bogoliubov transformation, that is, in terms of
a squeezed annihilation operator \citep{CNB-QOnotes} $\hat{c}=\hat{S}(r)\hat{a}\hat{S}^{\dagger}(r)=\hat{a}\cosh r-\hat{a}^{\dagger}\sinh r$,
with $\hat{S}(z)=\exp[(z\hat{a}^{\dagger2}-z^{*}\hat{a}^{2})/2]$,
so that (\ref{h}) takes the form
\begin{align}
\hat{h}\approx E_{0}(\alpha)+\Omega(\alpha)\hat{c}^{\dagger}\hat{c},
\end{align}
with\begin{subequations}
\begin{align}
\Omega & =\sqrt{\Delta^{2}-\sigma^{2}},\\
E_{0} & =(\Omega-\Delta)/2,\\
\sinh2r & =\sigma/\Omega.
\end{align}
\end{subequations}Note that stable, lower-bounded Hamiltonians require
$\sigma\geq0$ and $\Delta>\sigma$, which will be our case around
the classical minima. The ground state of this Hamiltonian is then
the vacuum state of the Bogoliubov mode $\hat{c}$, or, coming back
to the original mode and picture, the displaced squeezed vacuum state
$\hat{D}(\alpha)\hat{S}(r)|0\rangle$, with $\hat{a}|0\rangle=0$,
which has energy $E_{0}$. Let us now particularize these expressions
to the classical minima that we found above.

Consider first the $\varepsilon\leq1$ region, for which $\alpha=0$,
so that $\Delta=1$ and $\sigma=\varepsilon$, leading to $\Omega=\sqrt{1-\varepsilon^{2}}$
and $\sinh2r=\varepsilon/\Omega.$ In this case the approximate ground
state $\hat{S}(r)|0\rangle$ is unique and invariant under the $Z_{2}$
symmetry transformation. $\Omega$ provides the gap to the first excited
state, and closes at the critical point $\varepsilon=1$, where the
squeezing tends to infinity ($r\rightarrow\infty$).

When $\varepsilon>1$, two degenerate classical energy minima appear,
$\alpha=\pm\sqrt{L(\varepsilon-1)}\equiv\alpha_{\pm}$, so that around
either one of them $\Delta=2\varepsilon-1$ and $\sigma=1$, leading
to $\Omega=\sqrt{4\varepsilon(\varepsilon-1)}$ and $\sinh2r=1/\Omega.$
The degenerate ground space is in this case approximately spanned
by displaced squeezed vacua $\hat{D}(\alpha_{\pm})\hat{S}(r)|0\rangle$.
Both these states breaks the $Z_{2}$ symmetry spontaneously. The
ground-state energy can be written as $E_{0}=1/2-\varepsilon+\sqrt{(\varepsilon-1)\varepsilon}-L(\varepsilon-1)^{2}/2$.
Note that, once again, the squeezing diverges at the critical point.

The discussion above hints at the system possessing two phases, a
symmetry-preserving (disordered) and a symmetry-breaking (ordered)
phase, separated by the critical point $\varepsilon=1$. Note that
the fact that the squeezing, and hence the number of excitations $\langle\hat{a}^{\dagger}\hat{a}\rangle=\sinh^{2}r$,
diverges at that point is compatible with the mode exploring its underlaying
infinite-dimensional Hilbert space, as required for the existence
of a phase transition (in fact, we will later prove that $\langle\hat{a}^{\dagger}\hat{a}\rangle\sim L^{1/3}$
at the critical point, so that indeed $L$ controls how close we are
to the thermodynamic limit in this single-mode problem). However,
we emphasize that the Bogoliubov-de Gennes approach above does not
depend on $L$, is only approximate, and it diverges at the critical
point, making it hardly a proof that there exists a true phase transition
at the critical point. We then explore the problem with more accurate
techniques, including a better Gaussian ansatz and an approximation-free
numerical approach. Remarkably, the latter works for any value of
the parameters $(\varepsilon,L)$ by choosing an appropriate basis
of the Hilbert space as we explain now and detail in Appendices \ref{sec:Gaussian-minimization}
and \ref{sec:Numerical-Method}.

The numerical approach consists in finding the best Gaussian-state
\citep{CNB-QOnotes} estimate $\hat{D}(\bar{\alpha})\hat{S}(\bar{z})|0\rangle$
for the ground state of the system, and then building an appropriate
basis of the Hilbert space around it. The ground-state ansatz is found
by minimizing the energy functional $\langle\hat{H}\rangle$ with
respect to the complex parameters $\bar{\alpha}$ and $\bar{z}$.
We provide the details of this minimization in Appendix \ref{sec:Gaussian-minimization}.
Whenever $\bar{\alpha}=0$, we use the orthonormal set $\{\hat{S}(\bar{z})|n\rangle\}_{n=0,1,...,n_{\text{max}}}$
to represent and diagonalize the (sparse) Hamiltonian, where $|n\rangle$
are Fock eigenstates satisfying $\hat{a}^{\dagger}\hat{a}|n\rangle=n|n\rangle$,
and $n_{\text{max}}$ is a suitable truncation. If $\bar{\alpha}\neq0$
we then use a non-orthonormal set $\{\hat{D}(\pm\bar{\alpha})\hat{S}(\bar{z})|n\rangle\}_{n=0,1,...,n_{\text{max}}}$,
corresponding to basis vectors around the two degenerate minima. We
explain all the nuances related to the use of a non-orthonormal set
in Appendix \ref{sec:Numerical-Method}. Here it suffices to remark
that using this optimized basis, convergence is found for small values
of $n_{\text{max}}$ (say well below 100) no matter the choice of
$\varepsilon$ and $L$. This is because for small $L$ the number
of excitations $\langle\hat{a}^{\dagger}\hat{a}\rangle$ is small
as well, while for large $L$ the true ground state becomes closer
to the Gaussian state ansatz. In particular, we find that for large
but finite $L$ the non-Gaussian fluctuations have a non-negligible
(but still small) impact only extremely close to the critical point
$\varepsilon=1$. Moreover, for any $\varepsilon$ we can find a sufficiently
large value of $L$ for which the ground state becomes Gaussian for
all practical purposes. We illustrate this in Fig. \ref{Fig_GaussianOverlapFockDist}a,
where we plot the overlap between the true ground state and the Gaussian
approximation $\hat{S}(\bar{z})|0\rangle$ as a function of $L$ for
different values of $\varepsilon$ close to the critical point. For
any $\varepsilon$, the overlap is already very close to 1 for small
$L$, but jumps all the way to 1 (within numerical precision) as soon
as $L$ is large enough. From the tendency of the plot, we infer that
at the critical point $\varepsilon=1$ the overlap is exactly 1 only
in the thermodynamic limit $L\rightarrow\infty$. Nevertheless, in
Fig. \ref{Fig_GaussianOverlapFockDist}b we show that even for finite
$L$ the Gaussian ansatz $\hat{S}(\bar{z})|0\rangle$ contains almost
all the population of the ground state, with the contribution to any
other basis states $\hat{S}(\bar{z})|n\rangle$ decreasing exponentially
with $n$.

\begin{figure}[b]
\includegraphics[width=1\columnwidth]{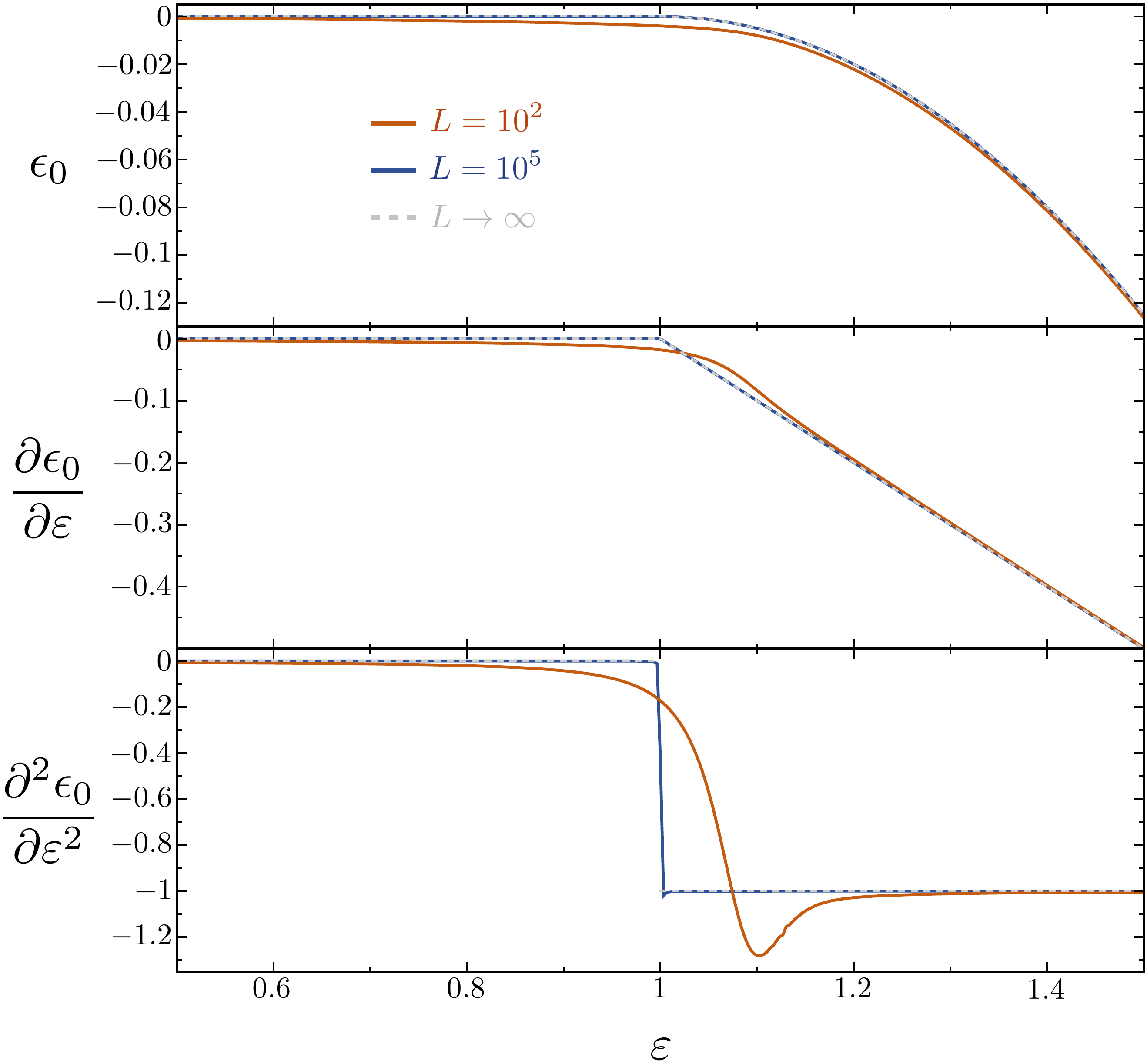}\caption{Ground-state energy density (and its derivatives) as a function of
the pair injection rate $\varepsilon$. The grey-dashed line is the
analytic prediction expected to be valid in the $L\rightarrow\infty$
limit, and predicting a discontinuity in the second order derivative
(second-order phase transition). The orange and blue lines are numerical
results for $L=10^{2}$ and $10^{5}$, respectively, which indeed
corroborate the expected analytical result. \label{Fig_E0PT}}
\end{figure}

\section{Main results}

\subsection{Phase transition and critical exponents\label{sec:Critical-Exponents}}

We now present the results that characterize the phase transition.
Let us start by discussing the ground-state energy density, which
according to the approximate results presented above is predicted
to be
\begin{equation}
\epsilon_{0}=\lim_{L\rightarrow\infty}\frac{E_{0}}{L}=\left\{ \begin{array}{cc}
0 & \text{for }\varepsilon\leq1\\
-(\varepsilon-1)^{2}/2 & \text{for }\varepsilon>1
\end{array}\right..\label{GroundStateEnergyVsSigma}
\end{equation}
Note first that this quantity is independent of $L$ (making the energy
extensive if $L$ is interpreted as sort of a system-size parameter
that controls how close we are to the thermodynamic limit, which we
prove throughout the next sections). We represent this quantity, as
well as its first and second order derivatives, as a function of $\varepsilon$
in Fig. \ref{Fig_E0PT}. Both $\epsilon_{0}$ and its first derivative
are continuous, while a discontinuity appears in the second derivative.
This approximate expression then predicts a second-order phase transition
in the thermodynamic limit. In the same figure we confront this prediction
with the numerical results found for increasing values of $L$. The
exact results approach the prediction of (\ref{GroundStateEnergyVsSigma})
as $L$ increases.

\begin{figure}[t]
\includegraphics[width=1\columnwidth]{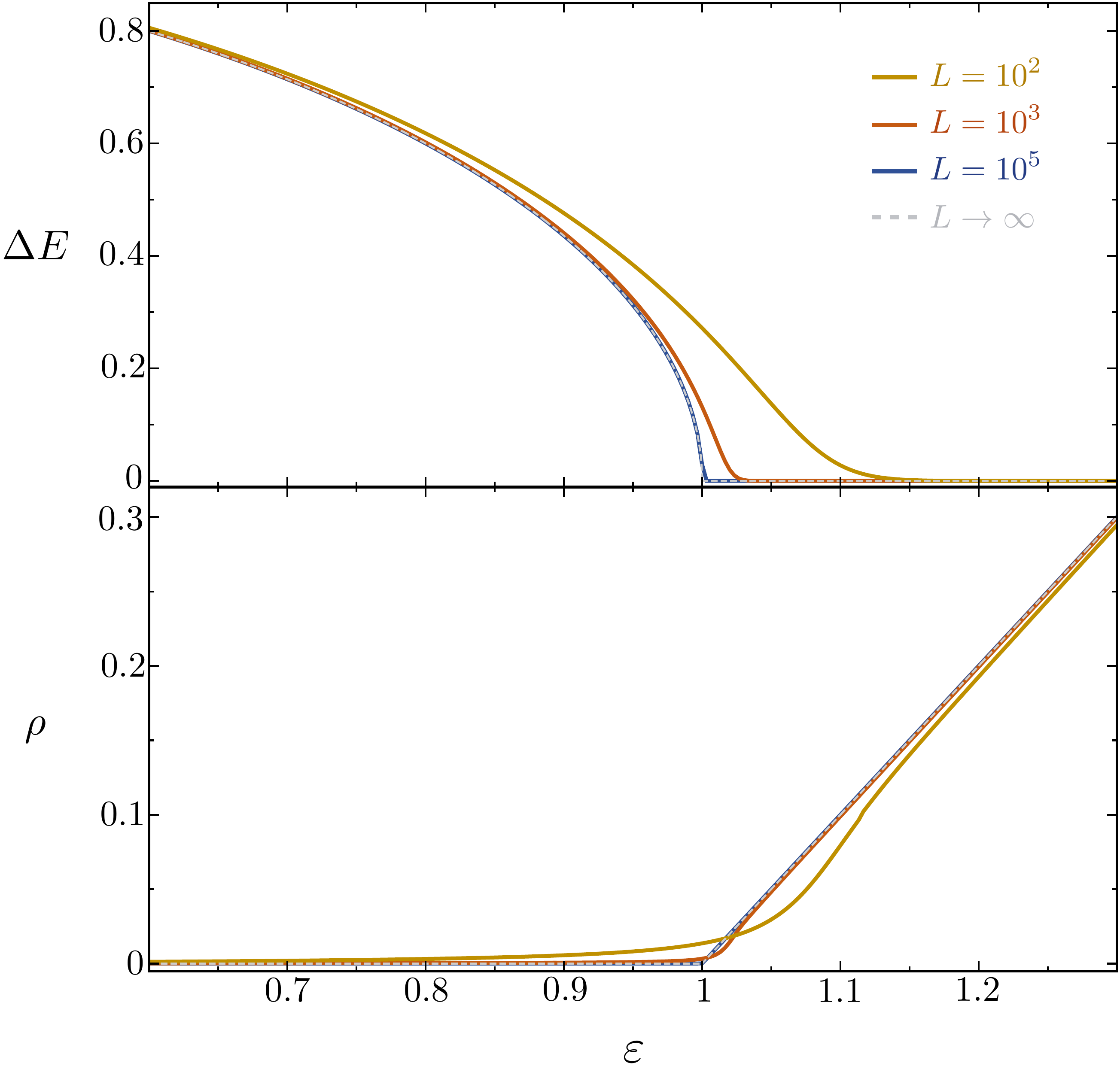}\caption{Energy difference $\Delta E$ between the two lowest eigen-energies
(top) and density of excitations $\rho$ (bottom) as a function of
the pair injection rate $\varepsilon$. Similarly to the previous
figure, the numerical results (solid lines) approach the expected
analytical results (dashed lines) in the $L\rightarrow\infty$ limit.
\label{Fig_GapOrderParameter}}
\end{figure}

In order to show that the phase transition belongs to the same universality
class as the one of the quantum Rabi and LMG models, we first analyze
the way in which observables behave when approaching the critical
point. We consider here two observables, the gap between the two smallest
Hamiltonian eigenvalues (counting degenerate ones as distinct), which
we denote by $\Delta E$, and the density of excitations $\rho=\langle\hat{a}^{\dagger}\hat{a}\rangle/L$.
In Fig. \ref{Fig_GapOrderParameter} we show these quantities as a
function of $\varepsilon$ for increasing values of $L$ (evaluated
numerically), together with the asymptotic results predicted by the
Bogoliubov approach of the previous section in the $L\rightarrow\infty$
limit:
\begin{align}
\Delta E & =\left\{ \begin{array}{cc}
\sqrt{1-\varepsilon^{2}} & \text{for }\varepsilon<1\\
0 & \text{for }\varepsilon>1
\end{array}\right.,
\end{align}
and
\begin{equation}
\rho=\left\{ \begin{array}{cc}
0 & \text{for }\varepsilon<1\\
\varepsilon-1 & \text{for }\varepsilon>1
\end{array}\right..
\end{equation}
The numerics confirm the Bogoliubov predictions. Moreover, given an
observable quantity $A$, we expect it to be characterized around
the critical point by a so-called critical exponent $\gamma_{A}$
via a power law 
\begin{equation}
\lim_{L\to\infty}A(\varepsilon,L)\propto\vert\varepsilon-1\vert^{\gamma_{A}}.\label{eq:criticalExponents}
\end{equation}
The Bogoliubov theory developed above provides us with the exponents
$\gamma_{\Delta E}=1/2$ and $\gamma_{\rho}=1$, the same exponents
as those found in the quantum Rabi and LMG models \citep{MyungJoongRabi}.
For completeness, it is also interesting to look at the uncertainty
of the position quadrature, $\hat{x}=\hat{a}^{\dagger}+\hat{a}$.
For a pure Gaussian state $\hat{S}(r)|0\rangle$, this is just given
by $\Delta x=e^{r}$. Noting that Bogoliubov-de-Gennes theory (below
the critical point) predicts $\sinh2r=\varepsilon/\sqrt{(1+\varepsilon)(1-\varepsilon)}$,
so $e^{2r}\propto1/\sqrt{1-\varepsilon}$ around the critical point,
we find $\gamma_{\Delta x}=-1/4$, exactly the one found in the quantum
Rabi model as well \citep{MyungJoongRabi}. 

\subsection{Finite-size scaling and exponents\label{sec:FiniteSizeScaling}}

\begin{figure}[b]
\centering{}\includegraphics[width=1\columnwidth]{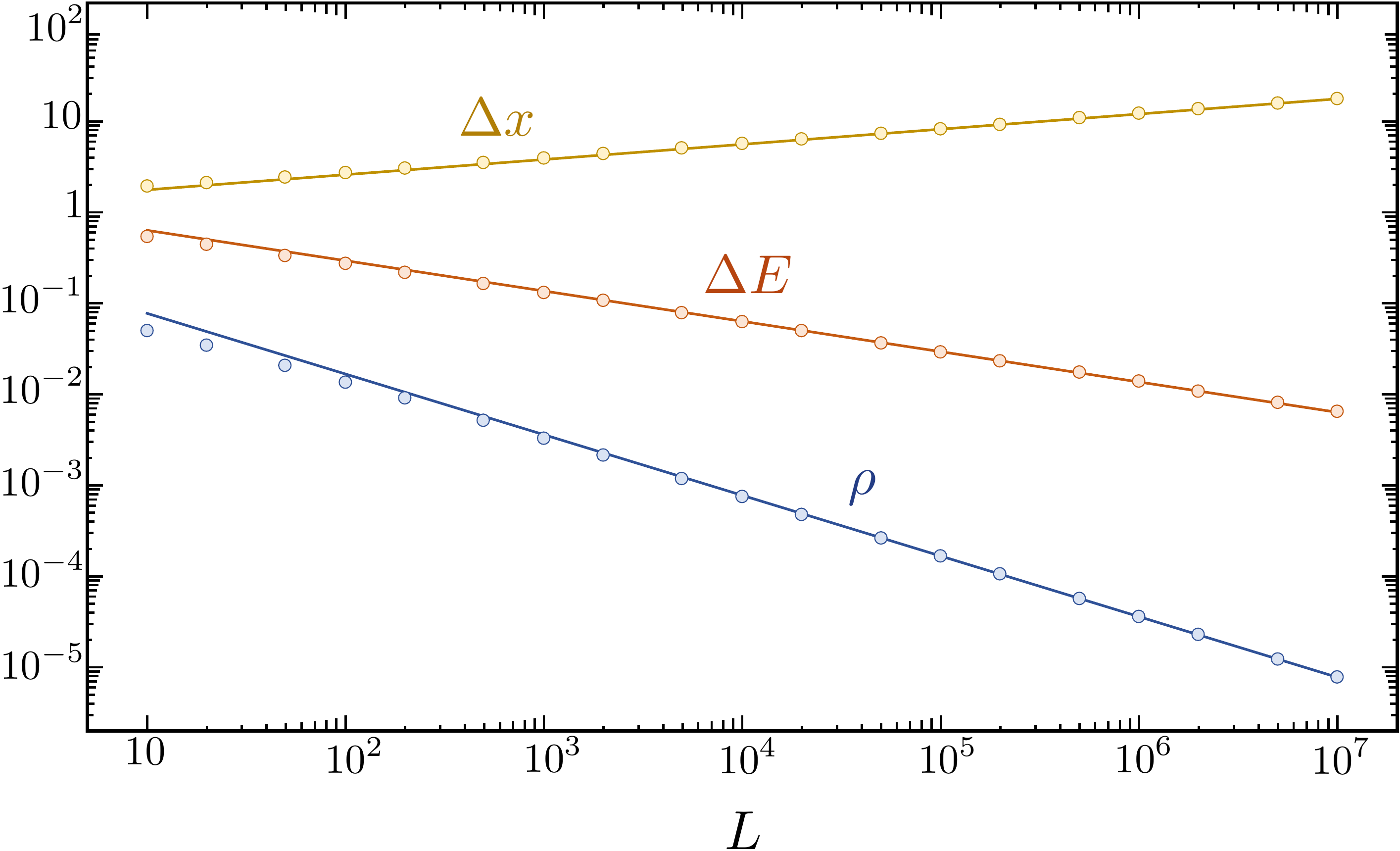}\caption{Log-log plot of the gap $\Delta E$ (red), density of excitations
$\rho$ (blue), and position uncertainty $\Delta x$ (yellow) as a
function of $L$ at the critical point $\varepsilon=1$. The circles
are numerical results, while the solid lines are the linear expressions
expected by the scalings found in the text. It is obvious that these
match perfectly as $L$ increases. \label{Fig_ScalingWithL}}
\end{figure}

In order to confirm that our single-mode model is in the same universality
class as the quantum Rabi and LMG models, we also need to analyze
how criticality is approached as $L$ increases. Here, we first consider
the finite-size exponent $\delta_{A}$ associated to an observable
$A$ of interest, defined through:
\begin{equation}
\lim_{\varepsilon\to1}A(\varepsilon,L)\propto L^{-\delta_{A}}.\label{eq:finiteSizeExponents}
\end{equation}
In Fig. \ref{Fig_ScalingWithL} we plot the same observables that
we considered in the previous section, $\Delta E$, $\rho$, and $\Delta x$,
but now as a function of $L$ for $\varepsilon=1$. The numerical
results show that they fit better and better a power-law of the type
(\ref{eq:finiteSizeExponents}) as $L$ increases. Bogoliubov theory
is not useful to determine the finite-size exponents, since it has
no information about $L$ (implicitly assumes $L\rightarrow\infty$
as explained above). However, we can still estimate them by considering
a squeezed-vacuum ansatz $\hat{S}(r)|0\rangle$, and minimizing the
energy functional $\langle0|\hat{S}^{\dagger}(r)\hat{H}\hat{S}(r)|0\rangle$
at $\varepsilon=1$. We show in Appendix \ref{sec:Gaussian-minimization}
that this provides a scaling relation $e^{2r}\propto L^{1/3}$, with
squeezing diverging as $L\rightarrow\infty$, consistently with the
Bogoliubov approach of the previous section. Using this ansatz, it
is then easy to find the finite-size exponents $\delta_{\Delta E}=-1/3$,
$\delta_{\rho}=-2/3$, and $\delta_{\Delta x}=1/6$ (see Appendix
\ref{sec:Gaussian-minimization} for details), which fit great the
numerical results, as shown in Fig. \ref{Fig_ScalingWithL}. Remarkably,
once again this scaling-law exponents coincide with those of the quantum
Rabi and LMG models \citep{MyungJoongRabi}.

\begin{figure}[t]
\includegraphics[width=1\columnwidth]{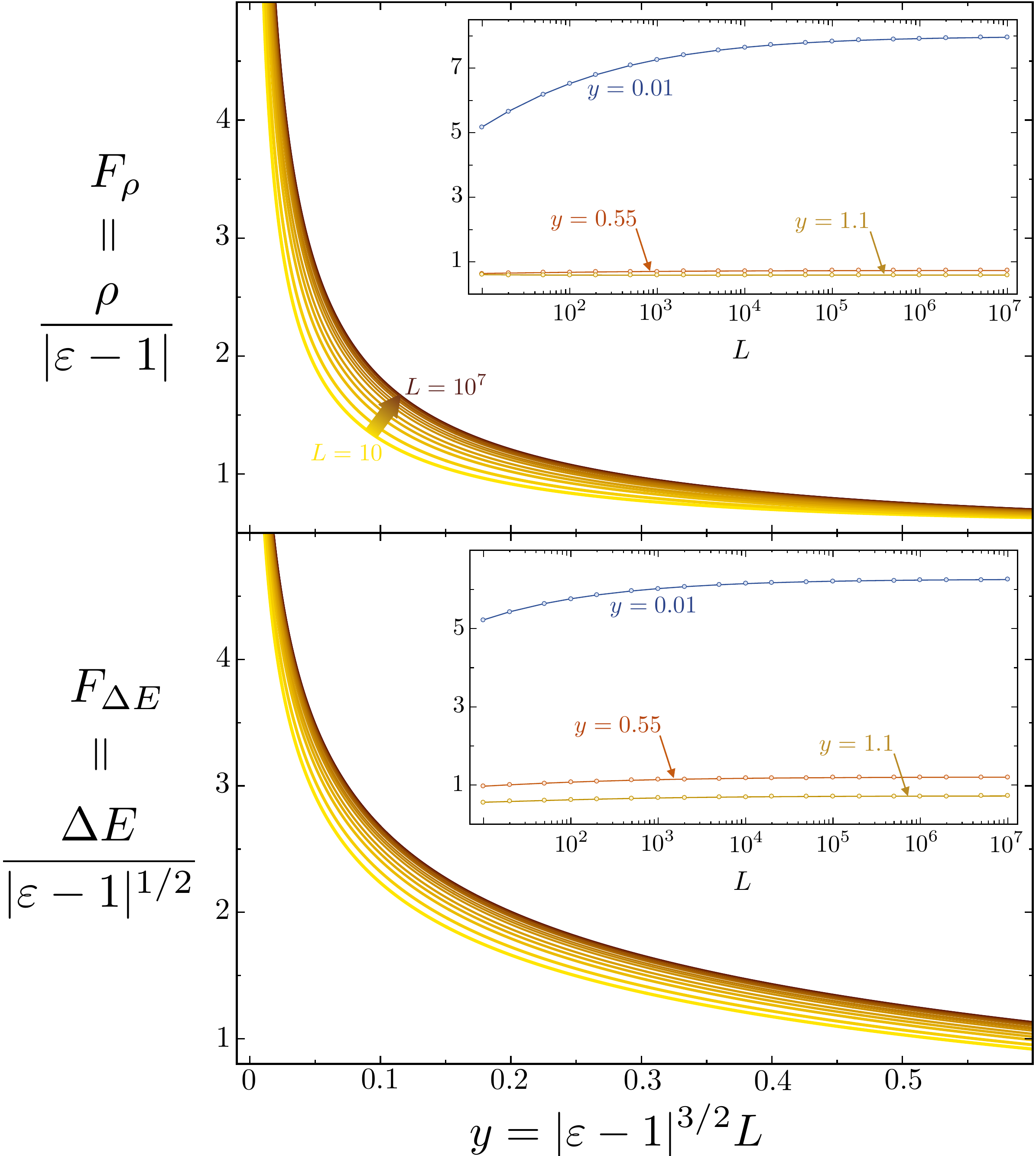}\caption{Scaling functions for the density of excitations $\rho$ (top) and
the gap $\Delta E$ (bottom). We can appreciate how the points obtained
numerically converge to a well-defined curve as $L$ increases (this
is most clear in the insets, which show that the scaling functions
saturate as a function of $L$ for fixed $y$). We remark that we
approach the critical point from below ($\varepsilon<1$). However,
we have checked that the same behavior appears when approaching it
from above. \label{Fig_ScalingFunctions}}
\end{figure}

\subsection{Scaling law\label{sec:ScalingLaw}}

In order to prove beyond any doubt that the model displays a second-order
phase transition and to completely determine its universality class,
we need to prove that all physical observables $A$ adhere to a scaling
law of the type \citep{KadanoffBook,GoldenfeldBook}
\begin{equation}
A(\varepsilon,L)=C_{1}|\varepsilon-1|^{\gamma_{A}}F_{A}(C_{2}|\varepsilon-1|^{\nu}L),\label{eq:scalingLaw}
\end{equation}
around the critical point $\varepsilon=1$ and the thermodynamic limit
$L\rightarrow\infty$. The so-called correlation-length exponent $\nu$
must be the same for all observables, while the function $F_{A}$
and the critical exponent $\gamma_{A}$ depend on the observable.
Models with the same scaling functions, critical exponents, and correlation-length
exponents are said to belong to the same universality class. The coefficients
$C_{j}$ can differ between models and observables. In loose terms,
the scaling law tells us that approaching the critical point either
by varying the control parameter $\varepsilon$ or the system's size
$L$ has the same effect, except for a well-defined scaling relation.
Note that expressions (\ref{eq:criticalExponents}) and (\ref{eq:finiteSizeExponents})
imply that the scaling function must satisfy\begin{subequations}\label{Fprops}
\begin{align}
\lim_{y\to\infty}F_{A}(y) & \propto1,\\
\lim_{y\to0}F_{A}(y) & \propto y^{-\frac{\gamma_{A}}{\nu}}.\label{cosa}
\end{align}
\end{subequations}The second line provides a relation between the
three characteristic exponents of the phase-transition, $\nu=\gamma_{A}/\delta_{A}$
for any observable $A$. For our model all three observables we have
considered lead to the same correlation-length exponent, $\nu=3/2$.

In Fig. \ref{Fig_ScalingFunctions} we show that our model satisfies
a scaling law (\ref{eq:scalingLaw}) by plotting $A(\varepsilon,L)|\varepsilon-1|^{-\gamma_{A}}$
as a function of $y=|\varepsilon-1|^{\nu}L$ for different values
of $\varepsilon$ and $L$, and two observables, the density of excitations
$\rho$ and the gap $\Delta E$. As $L$ goes towards infinity, all
the points converge towards a well-defined curve $F_{A}(y)$, that
satisfies the properties (\ref{Fprops}). Moreover, we have checked
that the scaling function $F_{A}$ is the same as the one of the quantum
Rabi model (after matching the scaling coefficients $C_{j}$ appropriately),
and hence we conclude that they are indeed in the same universality
class.

\begin{figure*}[t]
\centering{}\includegraphics[width=0.75\textwidth]{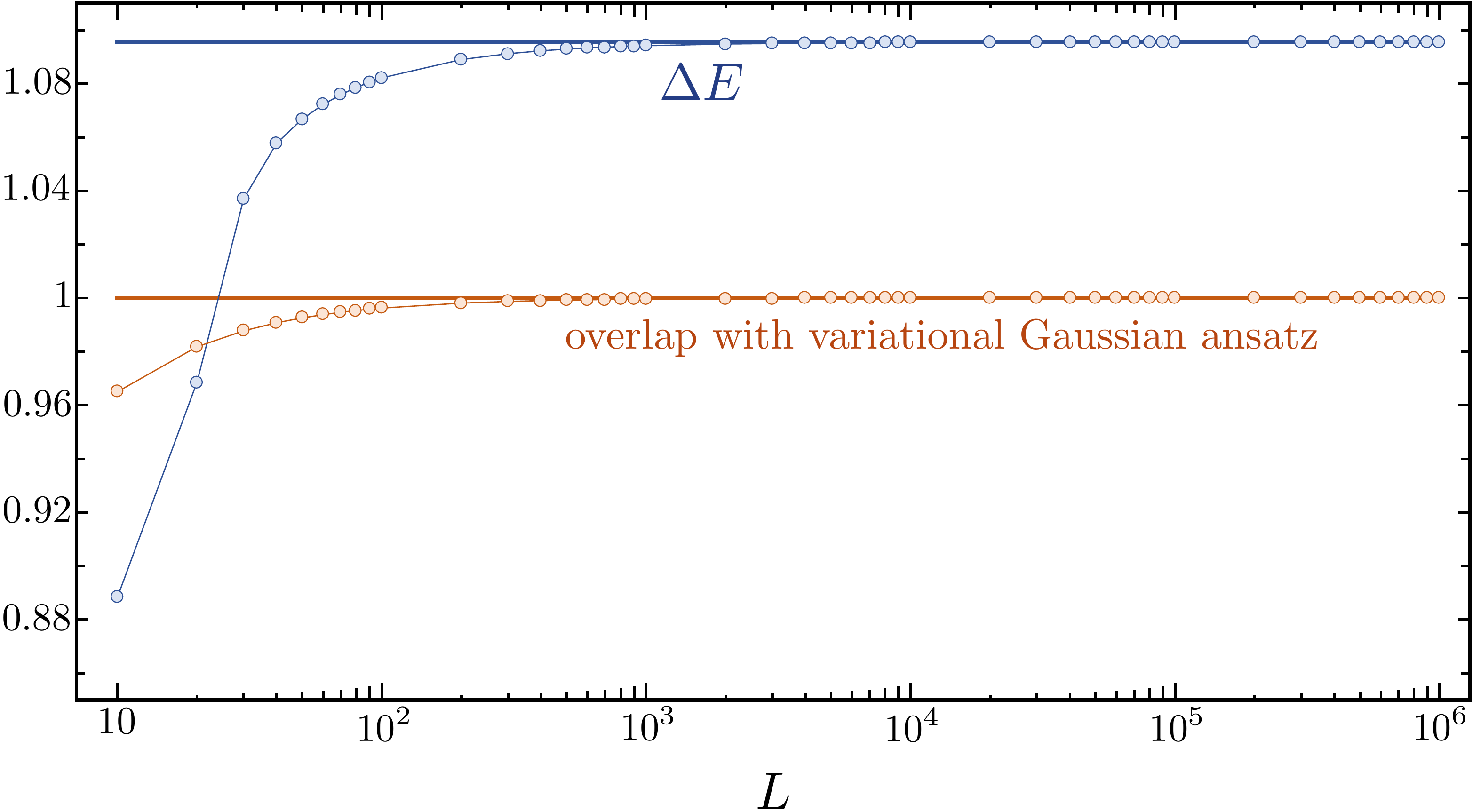}\caption{Gap between the two lowest energy eigenvalues (blue) and overlap of
the ground state with a variational displaced squeezed vacuum state
(red) as a function of $L$, when the Hamiltonian is perturbed by
a symmetry-breaking term with coupling strength $\lambda=1/\sqrt{L}$
for $\varepsilon=1.3$. The circles are numerical results, while the
solid lines are the results expected in the $L\rightarrow\infty$
limit. \label{Fig_Perturbation}}
\end{figure*}

\section{Sensitivity to symmetry-breaking perturbations\label{sec:Sensibility}}

An interesting feature of systems undergoing phase transitions and
spontaneous symmetry breaking is their sensitivity to infinitesimal
external perturbations while in the ordered phase. In our case, for
example, a finite gap should open in the ground-state manifold by
adding a symmetry-breaking perturbation of the type
\begin{equation}
\hat{V}=\frac{\mathrm{i}\lambda}{2}\left(\hat{a}-\hat{a}^{\dagger}\right),
\end{equation}
to the Hamiltonian (\ref{H}), with $\lambda\ll1$. By numerically
finding the two lowest-energy eigenstates, we show in Fig. \ref{Fig_Perturbation}
that, indeed, in the ordered region $\varepsilon>1$ a perturbation
$\lambda=1/\sqrt{L}$ is enough to open a gap in the ground-state
manifold of order larger than $\sqrt{\varepsilon-1}$.

We gain analytical insight by making use of perturbation theory. Specifically,
we consider the $L\to\infty$ limit, where the gap is closed for all
practical purposes in the absence of perturbation, so that the ground-state
manifold is two-dimensional, and approximately spanned by Gaussian
states $\hat{D}(\pm\bar{\alpha})\hat{S}(\bar{z})|0\rangle$ (see Appendix
\ref{sec:Gaussian-minimization} for details). In particular, the
dominant contribution is expected to come from the displacement with
$\bar{\alpha}=\sqrt{L(\varepsilon-1)}$, so that in the following
we approximate these states by coherent states $|\pm\bar{\alpha}\rangle$.
Furthermore, in this $L\rightarrow\infty$ regime these states can
be considered orthogonal, so that $\langle-\alpha|\hat{V}|\alpha\rangle=0$.
The corrections to the ground-state-manifold energies are then provided
by $\langle\pm\alpha|\hat{V}|\pm\alpha\rangle=\mp\lambda\alpha$ within
degenerate perturbation theory \citep{GriffithsBook}, hence predicting
the opening of a gap within the ground-state manifold given by
\begin{equation}
\Delta E=2\lambda\alpha=\sqrt{4\lambda^{2}L(\varepsilon-1)},\label{PerturbedGap}
\end{equation}
 which is finite in the ordered phase ($\varepsilon>1,L\rightarrow\infty$)
even for an infinitesimal parameter $\lambda\sim1/\sqrt{L}$.

In Fig. \ref{Fig_Perturbation} we plot the gap found numerically
for $\varepsilon=1.3$ and $\lambda=1/\sqrt{L}$, as a function of
$L$. We see that for large $L$ the numerical results converge to
expression (\ref{PerturbedGap}). In the same figure, we also plot
the overlap between the numerical ground state and the variational
Gaussian one $\hat{D}(\bar{\alpha})\hat{S}(\bar{z})|0\rangle$ (see
Appendix \ref{sec:Gaussian-minimization} for more details). We can
appreciate that these converge for large $L$, and moreover, we have
checked that the variational parameters converge to the ones that
we found with Bogoliubov theory in Section \ref{sec:Model-and-method}.

\section{Discussion and conclusions}

In summary, we have observed all the signatures required from a second-order
phase transition in a single-mode bosonic model. The phase transition
has been shown to be in the same universality class as those of more
complex non-extended models such as the quantum Rabi and LMG models,
making it the simplest model within this universality class (to our
knowledge). The simplicity of our model allows us to bring up a subtle
discussion related to whether one should call ``quantum'' phase transition
(meaning a phase transition in the equilibrium state of the system
at zero temperature, with the phase change driven by quantum fluctuations)
to the type found in non-extended systems. Let us elaborate on this.
First, let us point out that, as evidenced by the results above and
proven in more rigor in Appendix \ref{Appendix:Equivalence} for our
model, in these systems the thermodynamic limit coincides with the
classical limit, that is, with the limit in which quantum fluctuations
are irrelevant. In contrast, in extended systems the thermodynamic
limit is usually controlled by their size, and local quantum fluctuations
are still present even in the infinite-size limit. In this sense,
the phase transition of non-extended systems is less rich, and perhaps
one cannot even claim that it is robust against quantum fluctuations,
as they play no role in the thermodynamic or classical limit. Nevertheless,
it is our belief that such systems are still worth exploring, since
the elements required for a phase transition are present on them,
and being non-extended they can be used as building blocks of more
complex many-body models where the local phase transition of each
block competes with other types of phase transitions effected by many-body
phenomena.
\begin{acknowledgments}
We thank Myung-Joong Hwang, Xiangjun Xing, and Soonwon Choi for interesting
discussions. CNB thanks Valentina Hopekin for help with Figure \ref{Fig_GaussianOverlapFockDist},
and acknowledges sponsorship from the Yangyang Development Fund, as
well as support from a Shanghai talent program and from the Shanghai
Municipal Science and Technology Major Project (Grant No. 2019SHZDZX01).
\end{acknowledgments}

\appendix

\section{Gaussian minimization and finite-size exponents\label{sec:Gaussian-minimization}}

In this section we discuss the details of the Gaussian-state ansatz
$\hat{D}(\bar{\alpha})\hat{S}(\bar{z})\vert0\rangle$ that minimizes
the energy of the system for any value of $L$. This ansatz has been
crucial to determine the finite-size exponents in Section \ref{sec:FiniteSizeScaling},
to check the sensitivity to symmetry-breaking perturbations in Section
\ref{sec:Sensibility}, as well as to build a Hilbert-space basis
leading to efficient numerical simulations, as explained in Sections
\ref{sec:Model-and-method} and Appendix \ref{sec:Numerical-Method}
below. The variational parameters $\bar{\alpha}$ and $\bar{z}$ are
complex in principle, although we have found that the energy is minimized
when both are taken real with $\bar{z}$ positive. In particular,
the energy functional has the form
\begin{equation}
E(\bar{\alpha},\bar{z})=\langle\hat{a}^{\dagger}\hat{a}\rangle-\varepsilon\text{Re}\{\langle\hat{a}^{2}\rangle\}+\frac{\langle\hat{a}^{\dagger2}\hat{a}^{2}\rangle}{2L}.\label{Eappendix}
\end{equation}
We can use the identity \citep{CNB-QOnotes} $\hat{S}^{\dagger}(\bar{z})\hat{D}^{\dagger}(\bar{\alpha})\hat{a}\hat{D}(\bar{\alpha})\hat{S}(\bar{z})=\bar{\alpha}+\hat{a}\cosh\bar{r}+e^{\mathrm{i}\theta}\hat{a}^{\dagger}\sinh\bar{r}$,
with polar decomposition $\bar{z}=\bar{r}e^{\mathrm{i}\theta}$ for
the squeezing parameter, together with the definition $\delta\hat{a}\equiv\hat{a}-\langle\hat{a}\rangle$,
to write the following useful expressions: $\langle\hat{a}\rangle=\bar{\alpha}$,
$\langle\delta\hat{a}^{2}\rangle=e^{\mathrm{i}\theta}s_{2\bar{r}}/2$,
and $\langle\delta\hat{a}^{\dagger}\delta\hat{a}\rangle=s_{\bar{r}}^{2}$,
where we use the short-hand notation $s_{x}=\sinh x$. We then obtain
\begin{align}
\langle\hat{a}^{\dagger}\hat{a}\rangle & =|\bar{\alpha}|^{2}+s_{\bar{r}}^{2},\\
\langle\hat{a}^{2}\rangle & =\bar{\alpha}^{2}+e^{\mathrm{i}\theta}s_{2\bar{r}}/2,\nonumber \\
\langle\hat{a}^{\dagger2}\hat{a}^{2}\rangle & =|\bar{\alpha}|^{4}+4|\bar{\alpha}|^{2}s_{\bar{r}}^{2}+\tfrac{1}{4}s_{2\bar{r}}^{2}+2s_{\bar{r}}^{4}+\text{Re}\{\bar{\alpha}^{*2}e^{\mathrm{i}\theta}s_{2\bar{r}}\},\nonumber 
\end{align}
which lead to a relatively simple energy functional that is easily
minimized numerically with respect to the variational parameters with
the use of any mathematics software. Note that in the last equation
we have used the Gaussian moment theorem \citep{CNB-QOnotes} to write
$\langle\delta\hat{a}^{\dagger2}\delta\hat{a}^{2}\rangle=2\langle\delta\hat{a}^{\dagger}\delta\hat{a}\rangle^{2}+|\langle\delta\hat{a}^{2}\rangle|^{2}$,
removing odd moments of the fluctuation operators.

At the critical point $\varepsilon=1$ and for sufficiently large
$L$, we expect $\bar{\alpha}=0$ (which we also confirm numerically).
Under such circumstances, only the second term in (\ref{Eappendix})
depends on $\theta$, and is clearly minimized for $\theta=0$. The
energy functional can be written as a function of $x=e^{2\bar{r}}$
(expected to be large for large $L$), reading
\begin{equation}
E=\frac{3x^{2}}{32L}-\frac{x}{4L}-\frac{1}{2}+\frac{5}{16L}+\frac{1}{4x}\left(2-\frac{1}{L}\right)+\frac{3}{32Lx^{2}}.
\end{equation}
The derivative with respect to $x$ reads
\begin{equation}
\frac{dE}{dx}=\frac{3x}{16L}-\frac{1}{4L}-\frac{1}{4x^{2}}\left(2-\frac{1}{L}\right)-\frac{3}{16Lx^{3}},
\end{equation}
so that writing the minimization condition as 
\begin{equation}
4Lx^{3}\frac{dE}{dx}=\underbrace{\frac{3x^{4}}{4}-x^{3}-\frac{3}{4}}_{\approx3x^{4}/4}-\underbrace{(2L-1)}_{\approx2L}x=0,\label{MinCond}
\end{equation}
we obtain the scaling $x\sim L^{1/3}$ provided in Section \ref{sec:FiniteSizeScaling}.
Note that in (\ref{MinCond}) we have used $x\gg1$ and $L\gg1$,
but we have not assumed any particular dependence of $x$ with $L$.

Once we know the scaling of $e^{2\bar{r}}$, it is easy to find the
scaling for any observable. In the case of the number of excitations
$\langle\hat{a}^{\dagger}\hat{a}\rangle=s_{\bar{r}}^{2}\approx e^{2\bar{r}}/4$
and the quadrature variance $\langle\hat{x}^{2}\rangle=e^{2\bar{r}}$,
both of them scale with $L^{1/3}$, leading to the scalings provided
in Section \ref{sec:FiniteSizeScaling} for the density $\rho=\langle\hat{a}^{\dagger}\hat{a}\rangle/L$
and the uncertainty $\Delta x=\sqrt{\langle\hat{x}^{2}\rangle}$.

In order to determine the scaling of the gap, we take the variational
excited states $\hat{S}(\bar{r})|n\rangle$, for consistency with
our variational ground state. Using $\hat{S}^{\dagger}(\bar{r})\hat{a}\hat{S}(\bar{r})=\hat{a}\cosh\bar{r}+\hat{a}^{\dagger}\sinh\bar{r}$,
$\hat{a}|n\rangle=\sqrt{n}|n-1\rangle$, working at the critical point
$\varepsilon=1$, and defining the variational energy spectrum $E_{n}=\langle n|\hat{S}^{\dagger}(\bar{r})\hat{H}\hat{S}(\bar{r})|n\rangle$,
it is easy to show after some algebra that 
\begin{equation}
E_{1}-E_{0}=\frac{4\sinh^{4}r+2\sinh^{2}2r}{2L}+e^{-2\bar{r}}\sim L^{-1/3}.
\end{equation}

\section{numerical simulation\label{sec:Numerical-Method}}

In order to perform numerical diagonalization of the Hamiltonian,
it is crucial to choose an appropriate basis. Here we show how to
do this by building a basis from the Gaussian ansatz described in
the previous section. It is first convenient to rewrite the Hamiltonian
in terms of the Bogoliubov operators $\hat{c}=\hat{S}(\bar{r})\hat{a}\hat{S}^{\dagger}(\bar{r})$.
In particular, we simply use the relation $\hat{a}=\hat{c}\cosh r+\hat{c}^{\dagger}\sinh r$,
which inserted in (\ref{H}) leads to an expression for $\hat{H}$
as fourth-order polynomial in $\hat{c}$ and $\hat{c}^{\dagger}$.

Whenever the optimization provides $\bar{\alpha}=0$, we take $\{|n\rangle_{c}=\hat{S}(\bar{r})|n\rangle\}_{n=0,1,...,n_{\text{max}}}$
as an orthonormal basis, noting that $|n\rangle_{c}$ are the Fock
states associated to the Bogoliubov operators, that is $_{c}\langle m|\hat{c}|n\rangle_{c}=\sqrt{n}\delta_{m,n-1}$.
This leads to a simple sparse-matrix representation of $\hat{H}$,
whose eigenvector with smallest eigenvalue is efficiently found with
any algebra software, even when $n_{\text{max}}$ takes on extremely
large values. On the other hand, we have checked that for all our
simulations convergence is obtained way before that truncation, typically
well before $n_{\text{max}}\approx100$.

The situation is a bit more complex when $\bar{\alpha}\neq0$. In
such case, we take $\{\hat{D}(\pm\bar{\alpha})\vert n\rangle_{c}\}_{n=0,1,...,n_{\text{max}}}$
as the basis for the representation, built up from states localized
around the two possible $Z_{2}$-symmetry-breaking ground states.
The issue here is that states with opposite displacement are not orthogonal,
and hence, the basis vectors are not linearly independent. In order
to explain how we deal with this, let us define and sort the basis
elements as $\left\{ \vert\phi_{l}\rangle\right\} _{l=0,1,...,2n_{\text{max}}+1}$,
with even elements $|\phi_{2n}\rangle=\hat{D}(\bar{\alpha})|n\rangle_{c}$,
and odd ones $|\phi_{2n+1}\rangle=\hat{D}(-\bar{\alpha})|n\rangle_{c}$,
where $n=0,1,...,n_{\text{max}}$. We define the overlap matrix $\mathcal{A}$
with elements $\mathcal{A}_{kl}=\langle\phi_{k}\vert\phi_{l}\rangle\ne\delta_{kl}$.
We can insert the expansion $\vert\psi\rangle=\sum_{l}y_{l}\vert\phi_{l}\rangle$
into the eigenvalue equation $\hat{H}\vert\psi\rangle=\lambda\vert\psi\rangle$,
and apply $\langle\phi_{k}|$ onto it, obtaining the generalized eigenvalue
problem
\begin{equation}
\mathcal{H}\boldsymbol{y}=\lambda\mathcal{A}\boldsymbol{y},
\end{equation}
where the representation $\mathcal{H}$ of the Hamiltonian has components
$\mathcal{H}_{kl}=\langle\phi_{k}\vert\hat{H}\vert\phi_{l}\rangle$
and $\boldsymbol{y}=(y_{0},y_{1},...,y_{2n_{\text{max}}+1})^{T}$.
While this is no longer a sparse problem, the ground state can be
efficiently found with any algebra software as well, especially keeping
in mind that convergence is found for $n_{\text{max}}<100$ once again.
The only tricky point we want to emphasize is related to how to write
the matrices $\mathcal{A}$ and $\mathcal{H}$. In the case of the
overlap matrix, we use
\begin{align}
\mathcal{A}_{2q,2p} & =\langle\phi_{2q}\vert\phi_{2p}\rangle=\langle q\vert\hat{D}^{\dagger}(\bar{\alpha})\hat{D}(\bar{\alpha})\vert p\rangle=\delta_{qp},\\
\mathcal{A}_{2q,2p} & =\langle\phi_{2q+1}\vert\phi_{2p+1}\rangle=\langle q\vert\hat{D}^{\dagger}(-\bar{\alpha})\hat{D}(-\bar{\alpha})\vert p\rangle=\delta_{qp},\nonumber \\
\mathcal{A}_{2q+1,2p} & =\langle\phi_{2q+1}\vert\phi_{2p}\rangle=\langle q\vert\hat{D}^{\dagger}(-\bar{\alpha})\hat{D}(\bar{\alpha})\vert p\rangle\nonumber \\
 & =\langle q\vert\hat{D}(2\bar{\alpha})\vert p\rangle=\mathcal{D}_{qp}(2\bar{\alpha}),\nonumber \\
\mathcal{A}_{2q,2p+1} & =\langle\phi_{2q}\vert\phi_{2p+1}\rangle=\langle q\vert\hat{D}^{\dagger}(\bar{\alpha})\hat{D}(-\bar{\alpha})\vert p\rangle\nonumber \\
 & =\langle q\vert\hat{D}(-2\bar{\alpha})\vert p\rangle=\mathcal{D}_{qp}(-2\bar{\alpha}),\nonumber 
\end{align}
where $\mathcal{D}_{pq}(\alpha)={}_{c}\langle p|\hat{D}(\alpha)|q\rangle_{c}$
are the elements of the representation $\mathcal{D}(\alpha)$ of the
displacement operator $\hat{D}(\alpha)$ in the Bogoliubov mode's
Fock basis. Noting that the displacement operator can be written as
$\hat{D}(\alpha)=\exp(\alpha_{c}\hat{c}^{\dagger}-\alpha_{c}^{*}\hat{c})$,
with $\alpha_{c}=\alpha\cosh\bar{r}-\alpha^{*}\sinh\bar{r}$, we have
for $\alpha\in\mathbb{R}$ \citep{Cahill69}
\begin{equation}
\mathcal{D}_{pq}(\alpha)=e^{-|\alpha_{c}|^{2}/2}\times\left\{ \begin{array}{rr}
\sqrt{\frac{q!}{p!}}\alpha_{c}^{p-q}L_{q}^{(p-q)}(|\alpha_{c}|^{2}), & p\geq q\\
\sqrt{\frac{p!}{q!}}\alpha_{c}^{q-p}L_{p}^{(q-p)}(|\alpha_{c}|^{2}), & p<q
\end{array}\right.,
\end{equation}
where $L_{n}^{(a)}(x)$ are the generalized Laguerre polynomials.

Let us discuss now the representation of the Hamiltonian $\hat{H}$,
which we assume to be written in anti-normal order, such that it is
a sum of terms of the type $\hat{c}^{m}\hat{c}^{\dagger n}$ for different
natural values of $m$ and $n$. Note that, because the basis is not
orthonormal, the representation of a product of operators is no longer
the matrix product of their representations. Instead, we have to use
the following expressions:
\begin{align}
\langle\phi_{2q}\vert\hat{c}^{m}\hat{c}^{n\dagger}\vert\phi_{2p}\rangle & =\langle q\vert\hat{D}^{\dagger}(\bar{\alpha})\hat{c}^{m}\hat{D}(\bar{\alpha})\hat{D}^{\dagger}(\bar{\alpha})\hat{c}^{n\dagger}\hat{D}(\bar{\alpha})\vert p\rangle\nonumber \\
 & =\langle q\vert\left(\hat{c}+\bar{\alpha}_{c}\right)^{m}\left(\hat{c}^{\dagger}+\bar{\alpha}_{c}^{*}\right)^{n}\vert p\rangle\nonumber \\
=\sum_{r=0}^{m}\sum_{s=0}^{n} & \begin{pmatrix}m\\
r
\end{pmatrix}\begin{pmatrix}n\\
s
\end{pmatrix}\bar{\alpha}_{c}^{m-r}\left(\bar{\alpha}_{c}^{n-s}\right)^{*}\langle q\vert\hat{c}^{r}\hat{c}^{\dagger s}\vert p\rangle\nonumber \\
=\sum_{r=0}^{m}\sum_{s=0}^{n} & \begin{pmatrix}m\\
r
\end{pmatrix}\begin{pmatrix}n\\
s
\end{pmatrix}\bar{\alpha}_{c}^{m-r}\left(\bar{\alpha}_{c}^{n-s}\right)^{*}\\
 & \times\sqrt{\frac{(q+r)!(p+s)!}{q!p!}}\delta_{q+r,p+s},\nonumber 
\end{align}
with $\bar{\alpha}_{c}=\bar{\alpha}(\cosh\bar{r}-\sinh\bar{r})$,
the same for $\langle\phi_{2q+1}\vert\hat{c}^{m}\hat{c}^{n\dagger}\vert\phi_{2p+1}\rangle$
changing $\bar{\alpha}$ by $-\bar{\alpha}$, and
\begin{align}
\langle\phi_{2q+1}\vert\hat{c}^{m}\hat{c}^{n\dagger}\vert\phi_{2p}\rangle & =\langle q\vert\hat{D}^{\dagger}(-\bar{\alpha})\hat{c}^{m}\hat{c}^{n\dagger}\hat{D}(\bar{\alpha})\vert p\rangle\nonumber \\
 & =\langle q\vert\left(\hat{c}-\bar{\alpha}_{c}\right)^{m}\hat{D}(2\bar{\alpha})\left(\hat{c}^{\dagger}+\bar{\alpha}_{c}^{*}\right)^{n}\vert p\rangle\nonumber \\
=\sum_{r=0}^{m}\sum_{s=0}^{n}\begin{pmatrix}m\\
r
\end{pmatrix}\begin{pmatrix}n\\
s
\end{pmatrix} & (-\bar{\alpha}_{c})^{m-r}\left(\bar{\alpha}_{c}^{n-s}\right)^{*}\langle q\vert\hat{c}^{r}\hat{D}(2\bar{\alpha})\hat{c}^{\dagger s}\vert p\rangle\nonumber \\
=\sum_{r=0}^{m}\sum_{s=0}^{n}\begin{pmatrix}m\\
r
\end{pmatrix}\begin{pmatrix}n\\
s
\end{pmatrix} & (-\bar{\alpha}_{c})^{m-r}\left(\bar{\alpha}_{c}^{n-s}\right)^{*}\\
 & \times\sqrt{\frac{(q+r)!(p+s)!}{q!p!}}\mathcal{D}_{q+r,p+s}(2\bar{\alpha}),\nonumber 
\end{align}
plus the same for $\langle\phi_{2q}\vert\hat{c}^{m}\hat{c}^{n\dagger}\vert\phi_{2p+1}\rangle$
changing $\bar{\alpha}$ by $-\bar{\alpha}$. Note that in the second
equality, we have inserted the identity operator $\hat{D}(-\bar{\alpha})\hat{D}^{\dagger}(-\bar{\alpha})\hat{D}(\bar{\alpha})\hat{D}^{\dagger}(\bar{\alpha})$
in between $\hat{c}^{m}$ and $\hat{c}^{\dagger n}$.

\section{Equivalence between the thermodynamic and classical limits\label{Appendix:Equivalence}}

A neat way of showing that quantum fluctuations are fixed to zero-point
fluctuations in the thermodynamic limit $L\rightarrow\infty$ is by
using the positive $P$ phase-space representation of the state of
the system \citep{Drummond80,CarmichaelBook}, which allows mapping
the exact quantum dynamics into a set of stochastic differential equations.
The positive $P$ distribution can be seen as a generalization of
the more familiar Glauber-Sudarshan $P$ function $P(\alpha)$ \citep{CarmichaelBook1}.
The latter provides the coefficients required to expand a given density
operator $\hat{\rho}$ as a linear combination of coherent-state projectors,
that is, $\hat{\rho}=\int_{\mathbb{C}}\frac{d^{2}\alpha}{\pi}P(\alpha)|\alpha\rangle\langle\alpha|$.
Quantum expectation values in normal order are then mapped to phase-space
averages through $\langle\hat{a}^{\dagger m}\hat{a}^{n}\rangle=\int_{\mathbb{C}}\frac{d^{2}\alpha}{\pi}P(\alpha)\hat{\alpha}^{*m}\hat{\alpha}^{n}$.
Using the identities \citep{Drummond80,CarmichaelBook,CarmichaelBook1}
$\hat{a}|\alpha\rangle=\alpha|\alpha\rangle$ and $\hat{a}^{\dagger}|\alpha\rangle\langle\alpha|=(\alpha^{*}+\partial_{\alpha})|\alpha\rangle\langle\alpha|$,
it is easy to rewrite the von Neumann equation for the evolution of
the state, $\partial_{t}\hat{\rho}=-\mathrm{i}[\hat{H},\hat{\rho}]$,
as the following partial differential equation for the distribution
\citep{CarmichaelBook1}:
\begin{equation}
\partial_{t}P=\left[-\partial_{\alpha}A(\alpha)+\frac{1}{2}\partial_{\alpha}^{2}D(\alpha)\right]P+\text{c.c.},\label{FPeq}
\end{equation}
where\begin{subequations}
\begin{align}
A & =-\mathrm{i}(1+|\alpha|^{2}/L)\alpha+\mathrm{i}\alpha^{*},\\
D & =\mathrm{i}(\varepsilon-\alpha^{2}/L).
\end{align}
\end{subequations}In terms of the real variables $(\text{Re}\{\alpha\},\text{Im}\{\alpha\})$,
Eq. (\ref{FPeq}) has the form of a Fokker-Planck equation. However,
the corresponding diffusion matrix is easily shown not to be positive
semidefinite at some points of phase space. This is evidenced by the
naive application of the equivalence \citep{CarmichaelBook1} between
the Fokker-Planck equation and the following set of stochastic Langevin
equations: $\dot{\alpha}=A+\sqrt{D}\eta_{1}(t)$ and $\dot{\alpha}^{*}=A^{*}+\sqrt{D^{*}}\eta_{2}(t)$,
where $\eta_{j}(t)$ are independent real Gaussian white noises. These
are nonsensical equations, since $\alpha$ and $\alpha^{*}$ do not
remain complex-conjugate under evolution. The positive $P$ representation
$P_{+}(\alpha,\alpha^{+})$ is a generalization of the Glauber-Sudarshan
function that allows finding a Fokker-Planck equation with a positive
semidefinite diffusion matrix, but at the expense of working in a
doubled phase space, since here $\alpha$ and $\alpha^{+}$ are independent
complex variables. It can be (non-uniquely) defined through \citep{Drummond80,CarmichaelBook}
\begin{equation}
\hat{\rho}=\int_{\mathbb{C}^{2}}d^{2}\alpha d^{2}\alpha^{+}P_{+}(\alpha,\alpha^{+})\frac{|\alpha\rangle\langle\alpha^{+*}|}{\langle\alpha^{+*}|\alpha\rangle},
\end{equation}
for the representation of a state $\hat{\rho}$, where the states
in the kernel are coherent. It can be shown that quantum expectation
values are obtained as $\langle\hat{a}^{\dagger m}\hat{a}^{n}\rangle=\int_{\mathbb{C}^{2}}d^{2}\alpha d^{2}\alpha^{+}P_{+}(\alpha,\alpha^{+})\hat{\alpha}^{+m}\hat{\alpha}^{n}$.
In addition, it is also shown that the corresponding stochastic Langevin
equations can be obtained from the naive ones provided by the Glauber-Sudarshan
representation, just replacing $\alpha^{*}$ by $\alpha^{+}$ \citep{Drummond80,CarmichaelBook},
that is,\begin{subequations}\label{StochasticLangevinEqs}
\begin{align}
\dot{\alpha} & =-\mathrm{i}\left(1+\frac{\alpha^{+}\alpha}{L}\right)\alpha+\mathrm{i}\alpha^{+}+\sqrt{\mathrm{i}\left(\varepsilon-\frac{\alpha^{2}}{L}\right)}\eta_{1}(t),\\
\dot{\alpha}^{+} & =\mathrm{i}\left(1+\frac{\alpha^{+}\alpha}{L}\right)\alpha^{+}-\mathrm{i}\alpha+\sqrt{-\mathrm{i}\left(\varepsilon-\frac{\alpha^{+2}}{L}\right)}\eta_{2}(t).
\end{align}
\end{subequations}Since now $\alpha$ and $\alpha^{+}$ are independent
complex variables, there are no issues with them not evolving as a
complex-conjugate pair. It's important to note that these equations
provide the exact quantum dynamics of observables, there are no approximations
involved in their derivation. In addition, note that a coherent state
$|\alpha_{0}\rangle$ can be represented by the distribution $P_{+}=\delta^{(2)}(\alpha-\alpha_{0})\delta^{(2)}(\alpha^{+}-\alpha_{0}^{*})$,
meaning that whenever the solutions of (\ref{StochasticLangevinEqs})
do not fluctuate, the state of the system is given by a coherent state.
This is exactly what happens in the $L\rightarrow\infty$ limit as
we show next. In order to do this, let us normalize the variables
as $\alpha=\sqrt{L}\beta$ and $\alpha^{+}=\sqrt{L}\beta^{+}$, obtaining
then\begin{subequations}
\begin{align}
\dot{\beta} & =-\mathrm{i}\left(1+\beta^{+}\beta\right)\beta+\mathrm{i}\beta^{+}+\sqrt{\frac{\mathrm{i}}{L}\left(\varepsilon-\beta^{2}\right)}\eta_{1}(t),\\
\dot{\beta}^{+} & =\mathrm{i}\left(1+\beta^{+}\beta\right)\beta^{+}-\mathrm{i}\beta+\sqrt{-\frac{\mathrm{i}}{L}\left(\varepsilon-\beta^{+2}\right)}\eta_{2}(t).
\end{align}
\end{subequations}This normalized equations clearly show that in
the thermodynamic $L\rightarrow\infty$ limit, the noise term vanishes,
since $\beta$ and $\beta^{+}$ are finite. Hence, in this limit the
deterministic part of the equations dominates, and the variables do
not fluctuate when starting from non-fluctuating initial conditions.
In other words, a coherent state remains coherent. This shows that,
in our model, the thermodynamic and classical limits are not independent,
but equivalent.

\bibliographystyle{apsrev4-1}
\bibliography{ESMQPTBiblio}

\end{document}